# On the quantification of mixing in microfluidics

Ali Hashmi[a] and Jie Xu*[b]

[a] Mechanical Engineering, Washington State University, Vancouver, WA, USA.

[b] Mechanical Engineering, University of Illinois at Chicago, IL, USA. E-mail: jiexu@uic.edu

**Abstract**

Methods for quantifying mixing in microfluidics has largely varied in the past and various indices have been employed to represent the extent of mixing. Mixing between two or more colored liquids is usually quantified using simple mathematical functions operated over a sequence of images. The function, usually termed as mixing indices, involves a measure of standard deviation. Here we first review some mixing indices and then experimentally verify the index most representative of a mixing event. It is observed that the relative mixing index is not affected by the lighting conditions, unlike other known mixing indices. Based upon this finding, the use of relative mixing index is advocated for further usage in the lab on a chip community for quantifying mixing events.

**Introduction**

Mixing is one of the most fundamental challenges in microfluidics due to the laminar flow behaviour at microscale. Enhancing mixing in microfluidics has thus been extensively studied in past two decades, and being a quintessential function will continue to be of interest to the lab on a chip community. Generally speaking, mixing of fluids in microchannels can be achieved using either active or passive mixers[1-2]. Regardless of the mixing mechanism, there exists a need for a proper criterion to gauge the extent of mixing[3]. Usually, mixing is quantified by processing a set of images to yield a meaningful index that is representative of the extent of mixing. On the images, different fluids are usually differentiated based on differences in light intensities received by a camera. A dye is often used to absorb transmitted light, reflect incoming light or emit light itself for the camera recording. However, the extent of mixing remains incomparable throughout the wide spectrum of studies on microfluidic mixing. Indeed, none of the review papers have directly compared the mixing effects from different studies and a few are replete with misleading comparisons. Why we cannot compare mixing events across different studies, even if the researchers use the same mixing quantification method, is primarily due to the following reasons: 1) some mixing quantification methods are very sensitive to the initial mixing conditions - slight interfacial diffusion will influence the value dramatically; 2) some methods give values that depend on the minimum pixel intensity in the image – any mixing event that involves dyes other than black will be problematic; 3) in cases where mixing is not 1:1, some methods will produce values that are out of normal range.

Therefore, there is a need to set a common scale for quantifying the extent of mixing, making comparisons convenient across different studies. Here, we first summarize the various mixing indices and then establish a common index to gauge spatiotemporal mixing events more reliably.

**Mixing Indices**

Mixing indices are computed using intensities of pixels across a cross-section of a grayscale image that delineates a mixing event. The simplest index is calculated by taking the standard deviation, $\sigma$, of the pixel intensities[4-6], as shown in equation (1) below:

$$\sigma = \sqrt{\frac{1}{N}\sum_{i=1}^{N}(I_i - \langle I \rangle)^2} \qquad (1)$$

where $I_i$ shows a local pixel intensity, $\langle I \rangle$ shows the average of the pixel intensities in the cross-section, and N represents the total number of pixels.

This index attains the highest value when the fluids are unmixed and 0 if the fluids are homogeneously (completely) mixed. Although a measure of simple standard deviation yields information regarding the spread of data about the mean intensity, it is not a direct measure of the extent of mixing since the index possesses a dimension of intensity. Hence, comparison of mixing extent among different studies is not possible with the usage of this index.

The index can be rendered dimensionless by comparing the standard deviation to the mean intensity[7-16]. For the sake of discussion, we call the ratio as the "absolute mixing index (AMI)" and it can be computed from the following formula:

$$AMI = \frac{\sigma}{\langle I \rangle} = \frac{\sqrt{\frac{1}{N}\sum_{i=1}^{N}(I_i - \langle I \rangle)^2}}{\langle I \rangle} \qquad (2)$$

The aforementioned mixing index, on a scale of 0-1 (for 1:1 mixing), quantifies mixing where 1 delineates an unmixed state and 0 as a fully mixed state for the case in which one of the fluid streams yields a minimum pixel intensity of 0 (for instance, black ink that absorbs all the incoming light, or a black field without emission of fluorescent light). Note that, some researchers replace <I> with I_mixed in the equation, which serves the same purpose, since I_mixed should ideally equal to <I>.

Although AMI is a direct measure of the mixing extent, it still cannot be used for comparing mixing in different studies. For example, two hydrodynamically identical mixing events would result in different absolute mixing indices if different lighting conditions are used, let alone the cases where inks with dissimilar colors are used. Methods that tend to solve this problem include artificially rescaling (stretching) or normalizing each pixel to the same span of intensities as 0-255 on a grayscale image, or 0-1. The mixing index obtained using equations (2) with modified intensity values should be called "absolute mixing index with modified intensities".

The process of rescaling and stretching intensities makes analysis and comparison an extremely tedious process. Therefore, a third kind of mixing index, which we name here as "relative mixing index (RMI)" – also used in some earlier studies[17-20], precludes the need for any artificial pre-treatment of the intensity data, such as stretching or rescaling. RMI can be computed by taking the ratio of standard deviation of pixel intensities across a cross section to the standard deviation of the pixel intensities in the unmixed case, $\sigma_o$, according to the formula as follows:

$$RMI = 1 - \frac{\sigma}{\sigma_o} = 1 - \frac{\sqrt{\frac{1}{N}\sum_{i=1}^{N}(I_i - \langle I \rangle)^2}}{\sqrt{\frac{1}{N}\sum_{i=1}^{N}(I_{oi} - \langle I \rangle)^2}} \qquad (3)$$

where $I_{oi}$ represents the local pixel intensity in the unmixed state.

Note that, although the ratio itself is quite versatile in making comparisons of the extent of mixing, it is a rather non-intuitive representation. Thus, a more subtle way to quantify a mixing event is by considering (1–the ratio) in percentage form, which we can characterize as mixing efficiency or mixing ratio, so that the scale of RMI extends from 0 to 1, where 0 delineates the unmixed state and 1 as the homogeneously mixed state.

**Experimental**

Experiments were conducted to record a specific mixing event at varied light intensities, with images recorded using two different types of optical microscopes with different lighting direction. The extent of mixing was then compared using both AMI and RMI. A 7.92 x 10-4 Mol/L solution of blue dye (Erioglaucine disodium salt, Acros Organics) was allowed to mix with de ionized (D.I) water in a micro T-channel (Figure 1, left), which was fabricated using PDMS (Polydimethylsiloxane) soft-lithography.

Holes were punched into the polymer to create inlets and outlets for the passage of fluid. The chip was then plasma-bonded onto a glass slide. Inserts and tubing were attached to the two inlets of the microchannel on one end and to plastic syringes on the other to ensure fluid flow without any leakages. An insert with a tubing was attached to the outlet of the T-channel as a means to exhaust fluid. The syringes – one comprising of the blue dyed solution and the second with DI water - were mounted onto a syringe pump (KDS 210). The fluidic chip was placed under a Nikon microscope (MM400) attached to a camera (Nikon DS –Fi2 5 megapixel) to capture videos and images (10 X) in real time. The KDS pump was programmed to dispense fluids at 0.5 mL/min until a sharp interface was observed between the dyed solution and the DI water stream. Significant time was allowed for the system to reach equilibrium before the measurements were made. The image at time t=0 s was taken during the duration the pump was active. The pump was stopped and a stop watch was started; images were taken at various time intervals to capture the entire diffusion event (Figure 1, right). The experiments were performed for varied light intensities (labelled as bright, dark and darkest) using reflected light. Later, an experiment was performed under a stereomicroscope (Leica EZ4 HD) and images were captured at specified time intervals with transmitted light.

The images were analysed using a Matlab code to determine AMI and RMI. Results for the different cases are presented in Figure 2. As is evident, the AMIs deviate considerably for the same hydrodynamic mixing event at varied light conditions, whereas the computed RMI have an excellent agreement irrespective of the light intensity.

The caveat regarding the various mixing indices is that the entire characterization of mixing is based on the distribution pattern of the intensity values of the pixels within the span of the minimum and maximum intensities. On the one hand, standard deviation alone can be very different for similar patterns of intensity distribution. On the other hand, <I> is dependent on the light intensities and not necessarily in a linear fashion. Thus, usage of AMI is not an effective method of quantifying mixing events, except for cases when one of the fluids is absolutely black on the image. However, in RMI the temporally evolving distribution pattern of the pixel intensity values is always compared to the initial distribution. Hence, RMI as index is insensitive to variation in light intensities as well as the color of the dyes involved, provided the diffusion coefficient remains constant.

**Conclusions**

Since each biochemical event has its own characteristic time, therefore it is very important to have a clear comparison of mixing extent at different times scale. By quantifying one such mixing event under various light intensities and two different types of optical microscopes, we have demonstrated that AMI is not an accurate representative of the mixing extent. Owing to the insensitivity of RMI to light intensities, dye color, span of intensity distributions, we advocate its usage for future research in the scientific community.

## Acknowledgements

We thank the financial support from DARPA Young Faculty Award through grant N66001-11-1-4127.
## References

1. Lee, C. Y.; Chang, C. L.; Wang, Y. N.; Fu, L. M., Microfluidic Mixing: A Review. *Int J Mol Sci* **2011,** *12* (5), 3263-3287.
2. Nguyen, N. T.; Wu, Z. G., Micromixers - a review. *J Micromech Microeng* **2005,** *15* (2), R1-R16.
3. Danckwerts, P. V., The definition and measurement of some characteristics of mixtures. *Appl. sci. Res.* **1952,** *3* (4), 279-296.
4. Liu, R. H.; Stremler, M. A.; Sharp, K. V.; Olsen, M. G.; Santiago, J. G.; Adrian, R. J.; Aref, H.; Beebe, D. J., Passive mixing in a three-dimensional serpentine microchannel. *J Microelectromech S* **2000,** *9* (2), 190-197.
5. Stroock, A. D.; Dertinger, S. K. W.; Ajdari, A.; Mezic, I.; Stone, H. A.; Whitesides, G. M., Chaotic mixer for microchannels. *Science* **2002,** *295* (5555), 647-651.
6. Park, S. J.; Kim, J. K.; Park, J.; Chung, S.; Chung, C.; Chang, J. K., Rapid three-dimensional passive rotation micromixer using the breakup process. *J Micromech Microeng* **2004,** *14* (1), 6-14.
7. Stoeber, B.; Liepmann, D.; Muller, S. J., Strategy for active mixing in microdevices. *Physical Review E* **2007,** *75* (6), 066314.
8. Li, Y.; Zhang, D. L.; Feng, X. J.; Xu, Y. Z.; Liu, B. F., A microsecond microfluidic mixer for characterizing fast biochemical reactions. *Talanta* **2012,** *88*, 175-180.
9. Li, Y.; Xu, Y. Z.; Feng, X. J.; Liu, B. F., A Rapid Microfluidic Mixer for High-Viscosity Fluids To Track Ultrafast Early Folding Kinetics of G-Quadruplex under Molecular Crowding Conditions. *Anal Chem* **2012,** *84* (21), 9025-9032.
10. Garstecki, P.; Fischbach, M. A.; Whitesides, G. M., Design for mixing using bubbles in branched microfluidic channels. *Appl Phys Lett* **2005,** *86* (24).
11. Garstecki, P.; Fuerstman, M. J.; Fischbach, M. A.; Sia, S. K.; Whitesides, G. M., Mixing with bubbles: a practical technology for use with portable microfluidic devices. *Lab Chip* **2006,** *6* (2), 207-212.
12. Huang, P. H.; Xie, Y. L.; Ahmed, D.; Rufo, J.; Nama, N.; Chen, Y. C.; Chan, C. Y.; Huang, T. J., An acoustofluidic micromixer based on oscillating sidewall sharp-edges. *Lab Chip* **2013,** *13* (19), 3847-3852.
13. Mao, X. L.; Juluri, B. K.; Lapsley, M. I.; Stratton, Z. S.; Huang, T. J., Milliseconds microfluidic chaotic bubble mixer. *Microfluid Nanofluid* **2010,** *8* (1), 139-144.
14. Ahmed, D.; Mao, X. L.; Juluri, B. K.; Huang, T. J., A fast microfluidic mixer based on acoustically driven sidewall-trapped microbubbles. *Microfluid Nanofluid* **2009,** *7* (5), 727-731.
15. Venancio-Marques, A.; Barbaud, F.; Baigl, D., Microfluidic Mixing Triggered by an External LED Illumination. *J Am Chem Soc* **2013,** *135* (8), 3218-3223.
16. Lee, J. H.; Lee, K. H.; Won, J. M.; Rhee, K.; Chung, S. K., Mobile oscillating bubble actuated by AC-electrowetting-on-dielectric (EWOD) for microfluidic mixing enhancement. *Sensor Actuat a-Phys* **2012,** *182*, 153-162.
17. Lin, Y. C.; Chung, Y. C.; Wu, C. Y., Mixing enhancement of the passive microfluidic mixer with J-shaped baffles in the tee channel. *Biomed Microdevices* **2007,** *9* (2), 215-221.
18. Wang, S. S.; Huang, X. Y.; Yang, C., Mixing enhancement for high viscous fluids in a microfluidic chamber. *Lab Chip* **2011,** *11* (12), 2081-2087.
19. Xia, H. M.; Wang, Z. P.; Wang, W.; Fan, W.; Wijaya, A.; Wang, Z. F., Aeroelasticity-based fluid agitation for lab-on-chips. *Lab Chip* **2013,** *13* (8), 1619-1625.
20. Johnson, T. J.; Ross, D.; Locascio, L. E., Rapid microfluidic mixing. *Anal Chem* **2002,** *74* (1), 45-51.

**Figure Legends**

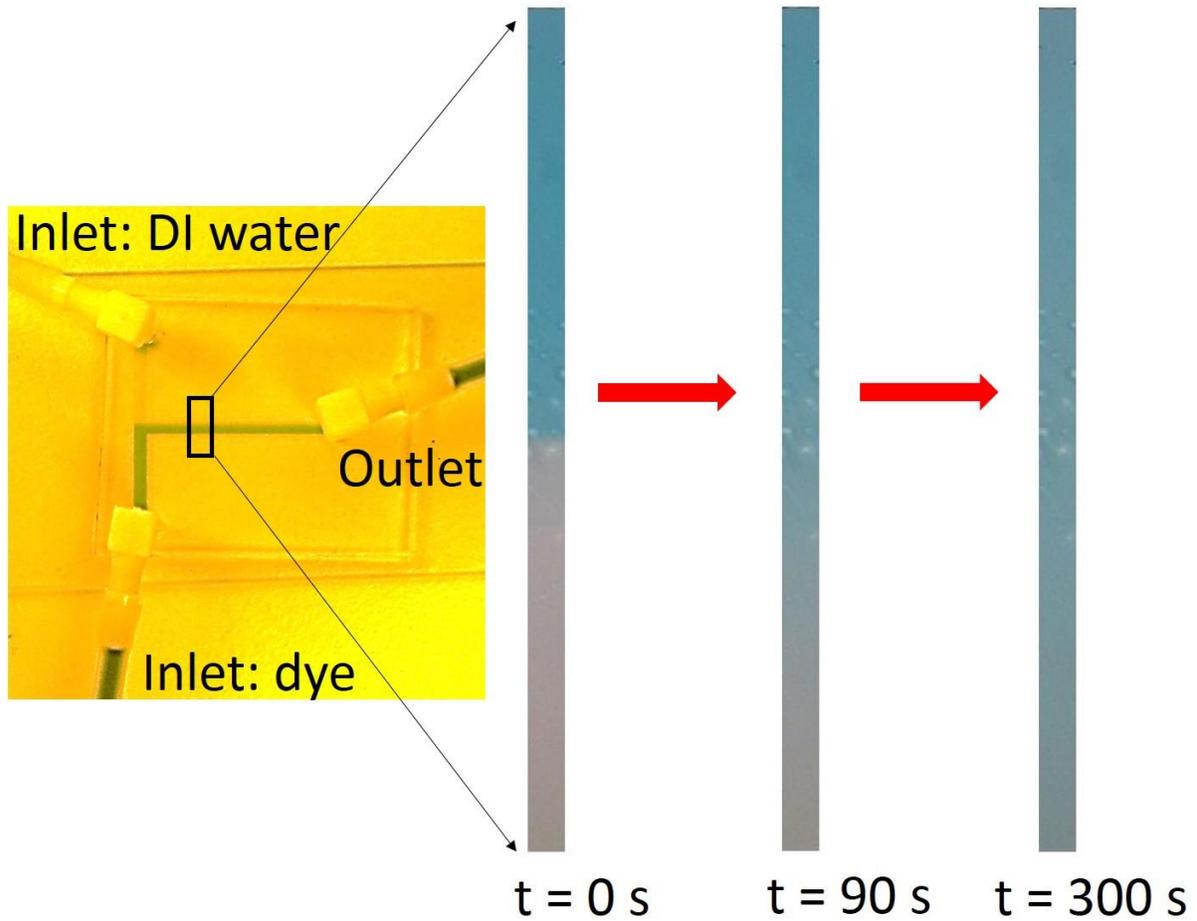

Figure 1: (left) a simple T-channel utilized for the experiments. (Right) the sequence of images shows the diffusion of blue dyed solution in water at three different time intervals, with the first image representing the unmixed case.

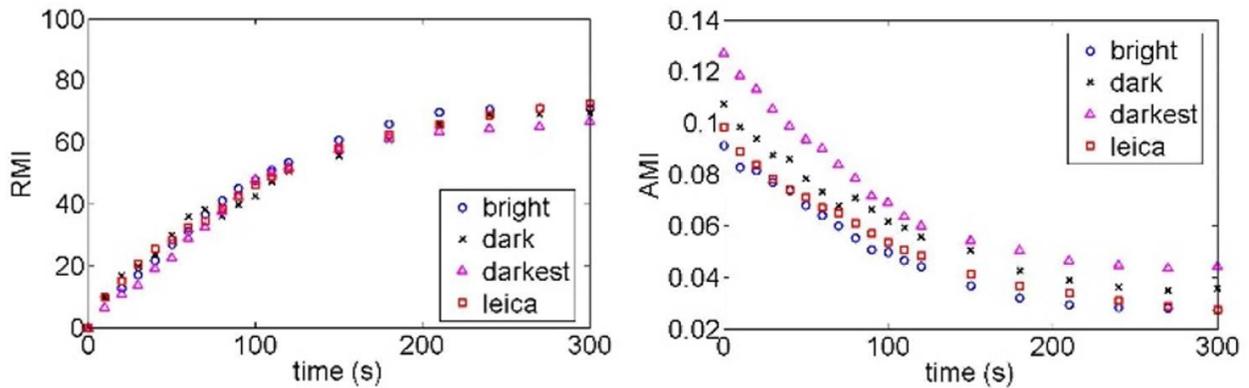

Figure 2: The graphs delineate the relative mixing index and absolute mixing index for the same mixing event at different light intensities under different types of microscopes. Significant deviation between absolute mixing indices and an excellent agreement between the relative mixing indices advocate the use of relative mixing index as a measure of the mixing extent.